\documentclass[11pt,a4paper]{article}
\usepackage[T1]{fontenc}
\usepackage[utf8]{inputenc}
\usepackage{lmodern}
\usepackage[UKenglish]{babel}
\usepackage[margin=1in]{geometry}
\usepackage{amsmath,amssymb,amsthm}
\usepackage{microtype}
\usepackage{url}
\usepackage[hidelinks,unicode]{hyperref}
\usepackage{aliascnt}

\newtheorem{theorem}{Theorem}
\newaliascnt{lemma}{theorem}
\newtheorem{lemma}[lemma]{Lemma}
\aliascntresetthe{lemma}

\newaliascnt{proposition}{theorem}
\newtheorem{proposition}[proposition]{Proposition}
\aliascntresetthe{proposition}

\addto\extrasUKenglish{%
}
\title{Destroying Rigidity in the Plane}
\author{%
  Yu Cong\thanks{Email: \href{mailto:yucong143@gmail.com}{\texttt{yucong143@gmail.com}}.}
  \qquad
  Chao Xu\thanks{Email: \href{mailto:thechaoxu@gmail.com}{\texttt{thechaoxu@gmail.com}}.}
  \qquad
  Yajie Zhao\thanks{Email: \href{mailto:yajiezhao8@gmail.com}{\texttt{yajiezhao8@gmail.com}}.}
  \\[0.7em]
  {\small University of Electronic Science and Technology of China, Chengdu, China}
}

\date{}
\hypersetup{
  pdftitle={Destroying Rigidity in the Plane},
  pdfauthor={Yu Cong, Chao Xu, Yajie Zhao},
  pdfsubject={Rigidity and global rigidity under edge and vertex deletion},
  pdfkeywords={rigidity, global rigidity, rigidity matroid, minimum cut}
}
\begin{document}
\maketitle
\begin{abstract}
Rigidity and global rigidity are fundamental graph properties in discrete
geometry, with applications to structural engineering and network
localization. In two dimensions, both admit combinatorial
characterizations in terms of the rigidity matroid and graph connectivity.
We study the computational complexity of finding minimum-cost sets of
edges or vertices whose deletion destroys either property in a finite
simple graph. With unit costs, we give deterministic polynomial-time
algorithms for edge deletion destroying global rigidity and vertex
deletion destroying either property. We prove that edge deletion destroying rigidity
is NP-hard and that, with positive integer costs, all four
problems are NP-hard. These results resolve, in dimension
two, the open questions on recognizing robustness under edge and vertex
deletions posed by Jordán and Whiteley \cite{JordanWhiteley2017}, as well
as the minimum edge-deletion problem for rigidity recorded in EGRES Open
\cite{EGRESDestroyingRigidity}.
\end{abstract}

\section{Introduction}\label{sec:introduction}

We study the computational complexity of finding minimum-cost edge or
vertex sets whose deletion destroys rigidity or global rigidity of a graph in the plane.
Rigidity theory studies whether distance constraints between vertices
are sufficient to determine the shape of a geometric framework.
In the plane, rigidity prevents nontrivial continuous deformations,
whereas global rigidity requires the entire realization to be unique
up to congruence \cite{AsimowRoth1978,Connelly2005}.
These notions arise naturally in structural engineering
\cite{GraverServatiusServatius1993}, sensor-network localization
\cite{AspnesEtAl2006}, and formation control \cite{YuAnderson2009},
where vertices represent joints, sensors, or agents, and edges
represent bars or prescribed distances. In practical settings,
however, distance constraints may be lost and individual vertices
may fail. This motivates studying not only whether a graph is rigid
or globally rigid, but also how robust these properties are under
edge and vertex deletions \cite{MotevallianYuAnderson2015,JordanWhiteley2017}.
In particular, we seek minimum-cost deletion sets that destroy
rigidity or global rigidity.
We distinguish rigidity of graphs from rigidity of frameworks.
Let $G=(V,E)$ be a simple graph. A \emph{placement} is a map
$p:V\to\mathbb R^2$, and the pair $(G,p)$ is a \emph{framework}.
For frameworks, rigidity asks whether every sufficiently small change in vertex
positions that preserves edge lengths also preserves all pairwise
distances; global rigidity asks the same question for arbitrary changes
in vertex positions.
Formally, a placement $q$ is \emph{equivalent} to $p$ if
$\|q(u)-q(v)\|=\|p(u)-p(v)\|$ for every edge $uv\in E$, and
\emph{congruent} to $p$ if this equality holds for all $u,v\in V$.
Congruent placements differ by an isometry of the plane. The framework
$(G,p)$ is \emph{rigid} if there is
$\varepsilon>0$ such that every placement $q$ equivalent to $p$
with $\|q(v)-p(v)\|<\varepsilon$ for every $v\in V$ is congruent to
$p$. It is \emph{globally rigid} if every placement equivalent to
$p$ is congruent to $p$.
Rigidity of a graph depends only on its graph structure.
A placement $p$ is \emph{generic} if its $2|V|$ coordinates are
algebraically independent over $\mathbb Q$. A graph $G$ is
\emph{rigid} or \emph{globally rigid} if $(G,p)$ has the corresponding
property for every generic placement $p$.
In two dimensions, rigidity and global rigidity of a graph can be determined through a matroid and combinatorial properties of the graph itself
\cite{AsimowRoth1978,Laman1970,Connelly2005,JacksonJordan2005}.
Throughout the paper, we use rigidity and global rigidity in this graph-theoretic sense.

\paragraph*{Previous work and open problems.}
Previous work has studied sufficient conditions and extremal bounds for
robustness under deletions. Motevallian, Yu and Anderson
\cite{MotevallianYuAnderson2015} study rigidity under multiple vertex
losses. A sufficient condition for global rigidity under such losses
follows from the theorem that every 6-connected graph is globally rigid
in the plane \cite[Theorem~7.2]{JacksonJordan2005}: a
$(k+5)$-connected graph remains 6-connected, and hence globally rigid,
after any deletion of at most $k-1$ vertices. Jordán determines
extremal edge counts for graphs that withstand every deletion of a given
number of edges or vertices while preserving rigidity or global rigidity,
for sufficiently large orders relative to the prescribed number of
deletions \cite{Jordan2021}.

The complexity of recognizing such robustness was explicitly posed as an
open problem by Jordán and Whiteley \cite[p.~1671]{JordanWhiteley2017}:
given a graph and an integer $k$, can one decide in polynomial time
whether every deletion of fewer than $k$ edges or vertices preserves
rigidity or global rigidity? These four recognition problems correspond
to testing whether the respective unit-cost minimum deletion value is at
least $k$. Enumerating all deletion sets of size less than $k$ gives
a polynomial-time algorithm for fixed $k$, but here $k$ is part of
the input. EGRES Open separately asks whether the minimum number of edges
whose deletion destroys rigidity in the plane can be computed in
polynomial time \cite{EGRESDestroyingRigidity}.

\paragraph*{Our results.}
We determine the complexity of all four deletion problems with unit costs
and with general costs, as summarized in \autoref{tab:results}.

\begin{table}[htbp]
\centering
\renewcommand{\arraystretch}{1.25}
\begin{tabular}{lcccc}
\hline
 & \multicolumn{2}{c}{Rigidity} & \multicolumn{2}{c}{Global rigidity} \\
Operation & Unit costs & General costs & Unit costs & General costs \\
\hline
Delete edges & NP-hard & NP-hard & P & NP-hard \\
Delete vertices & P & NP-hard & P & NP-hard \\
\hline
\end{tabular}
\caption{Complexity of deleting edges or vertices to destroy rigidity or global
rigidity. General costs are nonnegative real numbers. P means
deterministic polynomial time. All entries are proved here. General-cost
hardness holds already for positive integer costs on globally rigid
inputs. Unit-cost edge-deletion hardness holds on rigid inputs.}
\label{tab:results}
\end{table}

\begin{theorem}[Unit-cost deletion]
\label{thm:unit-cost-classification}
Given a finite simple graph $G$,
\begin{itemize}
\item computing the minimum edge subset whose deletion breaks rigidity of $G$ is NP-hard;
\item the minimum vertex subset whose deletion breaks rigidity of $G$ can be found deterministically in polynomial time;
\item the minimum edge or vertex subset whose deletion breaks global rigidity of $G$ can be found deterministically in polynomial time.
\end{itemize}
\end{theorem}

\autoref{thm:unit-cost-classification} resolves the recognition questions
of Jordán and Whiteley in dimension two \cite{JordanWhiteley2017}:
three admit deterministic polynomial-time algorithms, while deciding
whether every deletion of fewer than $k$ edges preserves rigidity is
coNP-hard. The edge-deletion hardness also settles the minimum-deletion
problem recorded in EGRES Open \cite{EGRESDestroyingRigidity}.

\begin{theorem}[Weighted deletion]
\label{thm:weighted-destruction}
Fix either edge deletion or vertex deletion, and either rigidity or
global rigidity. Given a globally rigid finite simple graph $G$,
positive integer costs on $E(G)$ or $V(G)$, respectively, taking at
most two distinct values, each at most $|V(G)|+|E(G)|+1$, and an
integer budget $b\geq0$, deciding whether a deletion of cost at most
$b$ destroys the chosen property is NP-hard.
\end{theorem}

\paragraph*{Organization.}

\autoref{sec:preliminaries} reviews the required rigidity and matroid
facts. \autoref{sec:algorithms} gives the three polynomial-time
algorithms. \autoref{sec:destruction} proves the hardness results for
weighted deletions and for unit-cost edge deletion destroying rigidity.

\section{Preliminaries}
\label{sec:preliminaries}

All graphs are finite and simple. For a graph $G=(V,E)$ and
$X\subseteq V$, we write $G[X]$ for the subgraph induced by $X$
and $G-X=G[V\setminus X]$. For $F\subseteq E$, the graph $G-F$
is obtained by deleting the edges in $F$. We write $d_G(v)$ for the
degree of a vertex $v$. A graph is \emph{$k$-connected} if it has
at least $k+1$ vertices and remains connected after deleting any set
of fewer than $k$ vertices. A \emph{vertex separator} of $G$ is a set
$Z\subseteq V$ for which $G-Z$ is disconnected.

We use standard matroid terminology \cite{Oxley2011}. Let $M$ be a
matroid on a finite ground set $E$, with rank function
$r_M:2^E\to\mathbb Z_{\geq0}$, where $r_M(F)$ is the maximum size of
an independent subset of $F$. A \emph{circuit} is a minimal dependent
set. An element $e$ is a \emph{coloop} if
$r_M(E\setminus\{e\})=r_M(E)-1$, that is, if $e$ lies in no circuit.
Call two elements related if they are equal or lie in a common circuit.
This is an equivalence relation, and its classes are the
\emph{components} of $M$. The rank of $M$ is the sum of the ranks of
its components.

The \emph{plane rigidity matroid} $\mathcal R_2(G)$ of a graph
$G=(V,E)$ has ground set $E$, and $F\subseteq E$ is independent if
and only if, for every $X\subseteq V$ with $|X|\geq2$, at most
$2|X|-3$ edges of $F$ have both endpoints in $X$.
Write $r_2(G)$ for the rank of $E$ in $\mathcal R_2(G)$.

\begin{theorem}[Laman's theorem, \cite{Laman1970}]
\label{thm:laman}
Let $G=(V,E)$ be a graph on $n\geq2$ vertices. Then
$r_2(G)\leq2n-3$, with equality if and only if $G$ is rigid.
Equivalently, $G$ is rigid if and only if it contains a spanning
subgraph with $2n-3$ edges satisfying the independence inequalities
above.
\end{theorem}

Graphs on at most one vertex are rigid. A graph is
\emph{redundantly rigid} if it is rigid and remains rigid after
deleting any edge. The \emph{support} of an edge set is the set of
endpoints of its edges.

\begin{lemma}[Components and circuits of the rigidity matroid, \cite{JacksonJordan2005}]
\label{lem:rigidity-matroid-components}
Let $G$ be a graph. Each singleton component of $\mathcal R_2(G)$
consists of a coloop. If a component $C$ with support $Z$ has at
least two edges, then $|Z|\geq4$ and $G[Z]$ is redundantly rigid.
Every component with support $Z$ has rank $2|Z|-3$.
Consequently, if $\ell(G)$ is the number of coloops and
$B_1,\ldots,B_t$ are the supports of the non-singleton components, then
\begin{equation}
\label{eq:component-rank}
r_2(G)=\ell(G)+\sum_{i=1}^t(2|B_i|-3).
\end{equation}
Moreover, if $C$ is a circuit with support $Z$, then the graph
$(Z,C)$ is redundantly rigid and has minimum degree at least three.
A rigid graph is redundantly rigid if and only if its rigidity
matroid has no coloops.
\end{lemma}

A \emph{rigid component} of $G$ is an inclusion-wise maximal vertex
set $R$ such that $G[R]$ is rigid.

\begin{lemma}[Connectivity and rigid components, \cite{JacksonJordan2005}]
\label{lem:rigid-components}
A rigid graph on at least three vertices is 2-connected.
Two rigid graphs sharing at least two vertices have a rigid union.
Consequently, distinct rigid components meet in at most one vertex,
and every rigid subgraph on at least two vertices lies in a unique
rigid component.
\end{lemma}

The connectivity assertion follows because otherwise one part of a
framework could be translated, or rotated about a cut vertex,
relative to the rest. Rigid components differ from matroid
components: a triangle is one rigid
component, but its three edges are singleton components of its rigidity
matroid.

\begin{theorem}[Global rigidity in the plane, \cite{Hendrickson1992,Connelly2005,JacksonJordan2005}]
\label{thm:global-rigidity}
A graph on at least four vertices is globally rigid if and only if
it is 3-connected and redundantly rigid. A graph on at most three
vertices is globally rigid if and only if it is complete.
\end{theorem}

We also use the following standard operations.

\begin{lemma}[Rank and vertex addition, {\cite{CruickshankEtAl2025}}]
\label{lem:rigidity-operations}
Adding an edge between two vertices of a rigid subgraph does not
change the rank of the ambient graph. Adding a new vertex adjacent
to exactly two distinct existing vertices increases the rank by two.
Consequently, adding a vertex with at least two distinct neighbors
preserves rigidity, and adding a vertex with at least three distinct
neighbors preserves redundant rigidity.
\end{lemma}

The first assertion follows from submodularity of rank, since a rigid
subgraph has the same rank as the complete graph on its vertex set.
The last assertion follows by applying the rigidity assertion after
each possible edge deletion.

\begin{lemma}[Substitution, {\cite[Lemma~2.12]{CruickshankEtAl2025}}]
\label{lem:rigidity-substitution}
Let $G_0,G_1,G_2$ be graphs with
$V(G_0)\cap V(G_1)=V(G_0)\cap V(G_2)=I$.
If $|I|\geq2$ and $G_0\cup G_1$ and $G_2$ are rigid, then
$G_0\cup G_2$ is rigid. The same implication holds for global
rigidity when $|I|\geq3$.
\end{lemma}

In particular, if $|I|\geq2$ and both $G_1$ and $G_2$ are rigid,
exchanging their roles shows that the two unions are rigid simultaneously. The analogous
equivalence holds for globally rigid $G_1,G_2$ when $|I|\geq3$.

The pebble game gives deterministic polynomial-time independence tests,
ranks, and bases in $\mathcal R_2(G)$
\cite{JacobsHendrickson1997,LeeStreinu2008}. \autoref{thm:laman} therefore
gives a deterministic polynomial-time rigidity test. Applying this test
to $G$ and to $G-e$ for every $e\in E$ checks redundant rigidity.
Combining these tests with a deterministic polynomial-time
3-connectivity test \cite{HopcroftTarjan1973} and the global-rigidity
characterization in \autoref{thm:global-rigidity} gives a deterministic
polynomial-time global rigidity test as well.

\section{Algorithms}
\label{sec:algorithms}

This section proves the polynomial-time part of
\autoref{thm:unit-cost-classification}. An input that already lacks the
target property needs no deletion, and both properties can be checked
deterministically in polynomial time. If no deletion set destroys the
target property, we assign the minimum deletion value $+\infty$ and
report infeasibility.

For edge deletion destroying global rigidity and vertex deletion
destroying rigidity, we apply rank additivity over the components of the
rigidity matroid (\autoref{lem:rigidity-matroid-components}) to the graph
left by an optimum deletion. In the first case, after excluding loss of
3-connectivity, it is rigid with a coloop in its rigidity matroid. In
the second case it is nonrigid. \autoref{lem:component-count} turns this
deficiency into a small edge cut or vertex separator of the original
graph. Vertex deletion destroying global rigidity then reduces to vertex
deletion destroying rigidity. Minimum edge cuts, including cuts
separating two specified vertex sets, and minimum vertex separators can
be found deterministically in polynomial time by standard network-flow
algorithms \cite{EdmondsKarp1972,Marx2006}.

\subsection{A component count}
\label{subsec:component-count}

For a graph $G$, use the coloop count $\ell(G)$ and the supports
$B_1,\ldots,B_t$ from \eqref{eq:component-rank}.
A vertex is \emph{uncovered}, \emph{private}, or \emph{shared} according
as it belongs to zero, one, or at least two of these supports. Let
$V_0$ be the set of uncovered vertices, and for each $i$, let
$s_i$ be the number of shared vertices in $B_i$. Define
$W(G)=2|V_0|+\sum_{s_i\leq2}(3-s_i)$. For each $i$, let $u_i$ be
the number of edges joining a private vertex of $B_i$ to a vertex
outside $B_i$.

\begin{lemma}
\label{lem:component-count}
Let $G$ be a simple graph on $n\geq2$ vertices. Then
$W(G)\geq2n-r_2(G)+\ell(G)$. Moreover,
$\sum_{v\in V_0}d_G(v)+\sum_{i:s_i\leq2}u_i\leq2\ell(G)$.
\end{lemma}

\begin{proof}
\label{proof:component-count}
Write $\ell=\ell(G)$. The expression
$2|V_0|+\sum_i(2|B_i|-s_i)$ counts each uncovered or
private vertex twice, and each shared vertex once for every support
containing it, hence at least twice. By \eqref{eq:component-rank},

\[
\begin{aligned}
2n&\leq2|V_0|+\sum_i(2|B_i|-s_i)\\
&=r_2(G)-\ell+2|V_0|+\sum_i(3-s_i)\\
&\leq r_2(G)-\ell+W(G).
\end{aligned}
\]

Every edge incident with an uncovered vertex or leaving a support at a
private vertex is a coloop, since it cannot belong to a non-singleton
component. Each endpoint contributes to at most one count. Thus each
coloop contributes at most two, proving the second inequality.
\end{proof}

Thus the rank deficiency of the rigidity matroid forces either many
uncovered vertices or components with few shared vertices, and these
structures will provide the small cuts and separators used in the
deletion algorithms below.

\subsection{Edge deletion destroying global rigidity}
\label{sec:global-deletion}

For a graph $G=(V,E)$ on $n\geq4$ vertices, let
$\operatorname{dg}(G)$ and $b_3(G)$ be the minimum numbers of edges
whose deletion destroys global rigidity and 3-connectivity, respectively.
For $X\subseteq V$, write $\delta_G(X)$ for the edges with exactly
one endpoint in $X$, and let
$
\lambda_2(G)=\min \{|\delta_G(X)|:2\leq|X|\leq n-2\}
$
be the minimum cut with at least two vertices on
each side. 

\begin{proposition}
\label{prop:global-cut-formula}
Let $G$ be a simple globally rigid graph on $n\geq4$ vertices.
If some single-edge deletion destroys global rigidity, then
$\operatorname{dg}(G)=1$. Otherwise,
$
\operatorname{dg}(G)=\min\{b_3(G),\lambda_2(G)-3\}
$.
\end{proposition}

\begin{proof}
\label{proof:global-cut-formula}
Assume every single-edge deletion preserves global rigidity.
By \autoref{thm:global-rigidity}, destroying 3-connectivity gives
$\operatorname{dg}(G)\leq b_3(G)$.
For a cut with at least two vertices on each side, redundant rigidity
requires at least $4$ crossing edges, since
\[
2n-3=r_2(G-e)\leq (2|X|-3)+(2|V\setminus X|-3)+|\delta_G(X)|-1
\]
for any $e\in\delta_G(X)$.
Deleting all but three destroys redundant rigidity, giving
$\operatorname{dg}(G)\leq\lambda_2(G)-3$.

For the reverse inequality, let $F$ be a minimum deletion destroying
global rigidity, and put $k=|F|\geq2$ and $H=G-F$. Suppose for a
contradiction that $k<b_3(G)$ and $k<\lambda_2(G)-3$.
Then $H$ is 3-connected. By minimality and
\autoref{thm:global-rigidity}, restoring any edge of $F$ gives a
redundantly rigid graph, so $H$ is rigid. Since $H$ is not globally
rigid, the same theorem and \autoref{lem:rigidity-matroid-components}
imply that it has $\ell=\ell(H)\geq1$ coloops.
Every non-singleton component support $B_i$ omits at least two vertices:
by \eqref{eq:component-rank}, a spanning support would force $\ell=0$,
while a support of size $n-1$ would leave just two coloops and no other
components, giving the remaining vertex degree at most two in $H$, in contradiction with the $3$-connectivity of $H$.

Use the uncovered vertices and supports of $H$, but let $u_i$ count
the edges of $G$ joining a private vertex
of $B_i$ to a vertex outside $B_i$. We have $d_G(v)\geq k+3$ for
every vertex, since reducing its degree to two destroys 3-connectivity.
If $s_i=0$, then $u_i=|\delta_G(B_i)|\geq k+4$.
If $s_i=1$ or $2$, deleting all but $2-s_i$ of the edges counted by
$u_i$ destroys 3-connectivity: remove the shared vertices and the
outside endpoints of the retained edges. Both sides remain nonempty,
as $|B_i|\geq4$ and $|V\setminus B_i|\geq2$.
Thus $u_i\geq k+3-s_i$ for $s_i=1,2$. Together with the bound for
$s_i=0$, this gives the uniform bound $u_i\geq2(3-s_i)+(k-2)$
whenever $s_i\leq2$.

By \autoref{lem:component-count}, $W(H)\geq\ell+3\geq4$.
In the sum defining $W(H)$, each uncovered vertex adds $2$, and each
support with $s_i\leq2$ adds $3-s_i\leq3$. Thus
$|V_0|+|\{i:s_i\leq2\}|\geq2$.
Applied to $H$, \autoref{lem:component-count} bounds the corresponding sum by $2\ell$,
and each edge of $F$ increases this sum by at most two. Hence, $\sum_{v\in V_0}d_G(v)+\sum_{i:s_i\leq2}u_i\leq 2\ell+2k$ in $G$.
Using the bounds above and $k\geq2$, we obtain
\[
\begin{aligned}
2\ell+2k
&\geq \sum_{v\in V_0}d_G(v)+\sum_{i:s_i\leq2}u_i\\
&\geq (k+3)|V_0|+\sum_{i:s_i\leq2}\bigl(2(3-s_i)+(k-2)\bigr)\\
&=2W(H)+(k-1)|V_0|+(k-2)|\{i:s_i\leq2\}|\\
&\geq 2W(H)+2(k-2)\\
&\geq 2(\ell+3)+2(k-2)
=2\ell+2k+2,
\end{aligned}
\]
a contradiction.
\end{proof}

Test every single-edge deletion and return one that destroys global
rigidity if it exists. Otherwise, compute $b_3(G)$ by enumerating
$Z\subseteq V(G)$ with $|Z|\leq2$ and finding a minimum edge cut of
$G-Z$. Deleting such a cut leaves $Z$ as a vertex separator, and
every edge deletion destroying 3-connectivity leaves a separator of size
at most two. Thus the smallest cut supplies $b_3(G)$ and a deletion
witness.
Every cut with at least two vertices on each side separates two disjoint
vertex pairs, and every cut separating two disjoint vertex pairs has at
least two vertices on each side. Hence $\lambda_2(G)$ and an attaining
cut are found by $O(n^4)$ minimum-cut computations, one for each
choice of two disjoint vertex pairs.
Contracting each prescribed vertex pair and replacing every undirected
edge by two opposite unit-capacity arcs gives a directed network with
$O(|E(G)|)$ arcs. The deterministic maximum-flow algorithm for
polynomially bounded integral capacities \cite{VanDenBrandEtAl2023}
therefore finds each such cut in $|E(G)|^{1+o(1)}$ time.
This proves the claim for edge deletion destroying global
rigidity in \autoref{thm:unit-cost-classification}.

Although \autoref{thm:global-rigidity} shows that an edge deletion destroying global rigidity must destroy either 3-connectivity or redundant rigidity, this characterization alone does not immediately yield a polynomial-time optimization algorithm. The cut characterization above supplies the additional structure needed here.

\subsection{Vertex deletion destroying rigidity}
\label{sec:vertex-deletion}

For $X\subseteq V(G)$, write
$N_G(X)=\{v\in V(G)\setminus X:\exists u\in X,\ uv\in E(G)\}$ for
its external neighborhood, and abbreviate $N_G(\{v\})$ to $N_G(v)$.
Write $\kappa(G)$ for the vertex connectivity.

We use separators to restrict the search for an optimum
vertex-deletion set to a polynomially enumerable family. For distinct
vertices $s,t$ of a graph $G$, an \emph{$s$--$t$ separator} is a
set $C\subseteq V(G)\setminus\{s,t\}$ such that $G-C$ has no
$s$--$t$ path. Let $\kappa_{s,t}(G)$ be its minimum size. For an
$s$--$t$ separator $C$, let $U(C)$ be the vertices reachable
from $s$ in $G-C$. The separator $C$ is \emph{important} if it is
inclusion-minimal and no $s$--$t$ separator $C'$ satisfies
$|C'|\leq|C|$ and $U(C')\supsetneq U(C)$ \cite{Marx2006}. Every
separator $C$ has an important replacement $C'$ with $|C'|\leq|C|$
and $U(C')\supseteq U(C)$: maximize the reachable set subject to these
bounds, then discard redundant separator vertices. Razgon's algorithm \cite[Theorem~3]{Razgon2010}
enumerates all important $s$--$t$ separators of size at most
$\kappa_{s,t}(G)+d$ deterministically in polynomial time for every
fixed integer $d\geq0$. Write $\rho_v(G)$ for the minimum number of
vertices whose deletion destroys rigidity.

\begin{proposition}
\label{prop:vertex-rigidity-deletion}
Let $G=(V,E)$ be a rigid graph. Then $\rho_v(G)$ can be found deterministically in polynomial
time.
\end{proposition}

\begin{proof}
If $G$ is a complete graph, vertex deletion never destroys its rigidity. Thus we assume that $G$ is not complete.
Write $\kappa=\kappa(G)$. Let $S$ be a minimum vertex separator, so
$|S|=\kappa$ and $G-S$ is disconnected.
Deleting all but one vertex of $S$ leaves a graph that is either
disconnected or has a cut vertex, and hence is nonrigid by
\autoref{lem:rigid-components}. Therefore
$1\leq\rho_v(G)\leq\kappa-1$. Keep this deletion set as the initial
candidate. It suffices to find an optimum of size at most $\kappa-2$
whenever one exists.

Let $X$ be such an optimum, put $k=|X|\leq\kappa-2$, and write
$H=G-X$. Then $H$ is 2-connected, since $\kappa(H)\geq\kappa-k\geq2$.
Suppose first that some vertex $v\in V(H)$ has degree at most five.
Since $d_G(v)\geq\kappa$, at most three vertices of $X$ lie outside
$N_G(v)$:
\[
|X\setminus N_G(v)|=k-d_G(v)+d_H(v)\leq3.
\]
We therefore enumerate a vertex $v$, at most five of its neighbors to
retain, and at most three vertices outside $N_G(v)\cup\{v\}$ to delete.
For each choice, delete all other neighbors of $v$ together with the
chosen additional vertices, and test whether this destroys rigidity.
Consider only deletions of size at most $\kappa-2$. This polynomial
search finds an optimum whenever some vertex of $H$ has degree at most
five.

Assume now that $H$ has minimum degree at least six.
Apply \autoref{lem:component-count} to $H$, and write $\ell=\ell(H)$.
Since $H$ is nonrigid, $W(H)\geq\ell+4$. If $u_i\geq6-2s_i$ whenever
$s_i\leq2$, that lemma gives the contradiction
\[
\begin{aligned}
2\ell
&\geq\sum_{v\in V_0}d_H(v)+\sum_{i:s_i\leq2}u_i\\
&\geq6|V_0|+\sum_{i:s_i\leq2}(6-2s_i)\\
&\geq2W(H)\geq2\ell+8.
\end{aligned}
\]
Fix a support $B=B_i$ with $s_i\leq2$ and $u_i\leq5-2s_i$.
By \autoref{lem:rigidity-matroid-components}, $H[B]$ is rigid and
$|B|\geq4$. Let $Z$ be the vertices of $B$ with neighbors in $V(H)\setminus B$.
Then $|Z|\leq s_i+u_i\leq5-s_i\leq5$.
Since $u_i\leq5-2s_i<2(|B|-s_i)$, some private vertex has at most one
neighbor outside $B$. Its degree is at most $|B|$, so $|B|\geq6$ and
$B\setminus Z$ is nonempty.

We next use $Z$ to replace part of $X$ by an important separator.
Let $U$ be a connected component of $H[B]-Z$. No edge joins $U$ to outside $B$
by the definition of $Z$.
Thus $N_H(U)\subseteq Z$. Also $B\ne V(H)$, since otherwise the rigid
graph $H[B]$ would equal $H$. Hence $Z$ is a separator of $H$, so
$2\leq |Z|\leq 5$. Put $Y=X\setminus N_G(U)$. Connectivity gives
\[\kappa\leq|N_G(U)|\leq |X\cap N_G(U)|+|Z|= k-|Y|+|Z|,\] hence
$|Y|\leq k-\kappa+|Z|\leq|Z|-2\leq3$.

We will keep $Y$ deleted and replace $X\setminus Y$ by an important
separator. Work in $G'=G-(Z\cup Y)$. Deleting
$X\setminus Y$ from this graph leaves $H-Z$, in which $U$ is a
connected component. Fix $s\in U$. Since $H$ is nonrigid, we can choose
$t\in V(H)$ outside the rigid component containing $B$.
Thus $X\setminus Y$ is an $s$--$t$ separator in $G'$ with reachable
set $U$. Let $C'$ be an important replacement, and let $U'$ be the
vertices reachable from $s$ in $G'-C'$. Then $U\subseteq U'$ and
$|C'|\leq|X\setminus Y|$, so $Y\cup C'$ deletes at most $k$ vertices.
We show deleting $Y\cup C'$ from $G$ destroys rigidity. Suppose instead
that $G-(Y\cup C')$ is rigid.
Every vertex of
$X\setminus(Y\cup C')$ has a neighbor in $U$, by the definition of
$Y$, and hence belongs to $U'$.
Set $I=B\setminus(C'\cup U')$ and define
$G_0=G-(Y\cup C'\cup U')$, $G_1=G[U'\cup I]$, and $G_2=H[B]$.
The preceding observation on $X\setminus(Y\cup C')\subseteq U'$ gives $G_0\subseteq H$.
We check the hypotheses of \autoref{lem:rigidity-substitution}.
Since $Z$ is disjoint from $Y\cup C'\cup U'$ and $Z\subseteq B$,
we have $Z\subseteq I$, so $|I|\geq|Z|\geq2$.
The definitions give
$V(G_0)\cap V(G_1)=V(G_0)\cap V(G_2)=I$.
Moreover, $U'$ is a connected component of $G'-C'$, so every
edge of $G-(Y\cup C')$ between $U'$ and $V(G_0)$ has its endpoint
in $V(G_0)$ in $Z\subseteq I$.
Hence $G_0\cup G_1=G-(Y\cup C')$, which is rigid by our supposition,
and $G_2=H[B]$ is rigid as established above.
The substitution lemma therefore implies that $G_0\cup G_2$ is rigid.
Since $G_0\subseteq H$ and $t\in V(G_0)$, this union is a rigid
subgraph of $H$ containing $B$ and $t$.
By \autoref{lem:rigid-components}, it lies in the unique rigid
component of $H$ containing $B$, contrary to the choice of $t$.
Thus deleting $Y\cup C'$ destroys rigidity.

We can now enumerate the candidate deletion sets.
Enumerate $Z$ with $2\leq|Z|\leq5$, then disjoint $Y$ with
$|Y|\leq|Z|-2$. In $G'=G-(Z\cup Y)$, enumerate nonadjacent vertices
$s,t$ and all important $s$--$t$ separators $C'$ of size at most
$\kappa-2-|Y|$, since we only need to find an optimum with $\rho_v(G)\leq\kappa-2$ in this case. Adding $Z\cup Y$ to any such separator gives a
separator of $G$, so
$\kappa_{s,t}(G')\geq\kappa-|Z|-|Y|$.
Thus the size bound exceeds the minimum $s$--$t$ separator size
$\kappa_{s,t}(G')$ by at most $|Z|-2\leq3$.
Then Razgon's algorithm \cite[Theorem~3]{Razgon2010} enumerates the separators in polynomial time.
Test each $Y\cup C'$ for destruction of rigidity and return the
smallest successful set.
This gives a deterministic polynomial-time algorithm for $\rho_v(G)$.
\end{proof}

\subsection{Vertex deletion destroying global rigidity}
\label{sec:global-vertex-deletion}

Recall that we write $\rho_v(G)$ for the minimum number of
vertices whose deletion destroys rigidity. Vertex deletion can never destroy
global rigidity in complete graphs, so we exclude them.
\begin{proposition}
\label{prop:global-vertex-deletion}
Let $G$ be a globally rigid noncomplete graph on $n$ vertices. Then
the minimum number of vertices whose deletion destroys global rigidity
is the minimum of $\kappa(G)-2$ and the values $\rho_v(G-e)$ over
all $e\in E(G)$. An optimum deletion set can be found
deterministically in polynomial time.
\end{proposition}

\begin{proof}
\label{proof:global-vertex-deletion}
Put $q=\kappa(G)-2$. Deleting all but two vertices of a minimum
separator supplies a solution of size $q$, since the resulting graph is only $2$-connected.
We have $q\leq n-4$ whenever $G$ is not a complete graph. Every deletion set $X$ with $|X|<q$ therefore leaves a
3-connected graph on at least four vertices. By
\autoref{thm:global-rigidity},
\[
G-X\text{ is not globally rigid}
\quad\Longleftrightarrow\quad
(G-e)-X\text{ is nonrigid for some }e\in E(G).
\]
Together with the solution of size $q$, this proves the formula.
Each $G-e$ is rigid and noncomplete, so
$\rho_v(G-e)$ and a witness can be computed deterministically in polynomial time by \autoref{prop:vertex-rigidity-deletion}. Return the separator-based solution unless
one of these witnesses has size less than $q$, in which case return a
smallest such witness.
\end{proof}

\section{Hardness}
\label{sec:destruction}

All hardness results reduce from minimum $k$-cut on connected simple
graphs, which asks for a minimum edge set whose deletion leaves at least
$k$ components. It is polynomial-time solvable for fixed $k$
\cite{GoldschmidtHochbaum1994} but NP-hard when $k$ is part of the
input, even for connected simple graphs, by the clique reduction in
\cite{Li2019}. The reductions need graphs
whose rigidity and global rigidity are controlled by the number of
connected components of a source graph $H$, so that cheap deletions
correspond to edge deletions in $H$.

\subsection{The tree construction}
\label{sec:tree-construction}

Let $\gamma(G)$ be the number of connected components in graph $G$,
including isolated vertices. Let $H=(V,E)$ be a simple graph with
$V=\{v_1,\ldots,v_N\}$, where $N\geq4$. We call these vertices terminals.
Let $h\geq0$ be an integer. Fix a binary tree $T_N$ whose
leaves are exactly the terminals and whose internal vertices all have
degree three. It has $N-2$ internal vertices and $2N-3$ edges.
We construct $Q_h(H)$ as follows. Start with the graph $H$ on the terminal set $\{v_1,\ldots,v_N\}$,
and add a new vertex $a$ together with the edges $av_i$ for all $i=1,\ldots,N$. Next, for each $j=1,\ldots,h$,
take a copy $T_N^{(j)}$ of $T_N$ and identify its leaf labeled $v_i$ with
the corresponding terminal $v_i$ of $H$. All internal vertices of $T_N^{(j)}$ are newly introduced and are
distinct from $a$ and other copies. Add all edges of each tree copy, while retaining all original edges of $H$.

\begin{lemma}
\label{lem:tree-rigidity}
Let $Q_h(H)$ be given by the construction above, and put
$n=|V(Q_h(H))|$ and $\gamma=\gamma(H)$.
Then $r_2(Q_h(H))=\min\{2n-3,2n-2-\gamma+h\}$.
If $h\geq1$ and $f$ is any edge in a tree copy, then
$r_2(Q_h(H)-f)=\min\{2n-3,2n-3-\gamma+h\}$.
\end{lemma}

\begin{proof}
\label{proof:tree-rigidity}
We use the rank and vertex-addition facts in
\autoref{lem:rigidity-operations}.

Let $F$ be a spanning forest of $H$, obtained by choosing a spanning tree in each connected component of $H$.
We first show that replacing $H$ by $F$ does not change the rigidity rank of the constructed graph. Consider any
component $C$ of $F$, and choose a root $r$ of its tree. We start with the edge $ar$ and add the remaining vertices of
$C$ in an order such that each vertex is added after its parent. When a vertex $v$ is added, connect it to both $a$ and
its parent in $C$. Since these are two distinct vertices already present, each addition preserves rigidity by
\autoref{lem:rigidity-operations}. Thus the resulting subgraph is rigid. By the rank property in \autoref{lem:rigidity-operations},
adding an edge between two vertices of a rigid subgraph does not increase the rigidity rank of the ambient graph.
Therefore, restoring all edges of $E(H)\setminus E(F)$ to $Q_h(F)$ does not change its rank. Since this produces exactly $Q_h(H)$,
we obtain $r_2(Q_h(H))=r_2(Q_h(F))$. Consequently, it suffices to work with $Q_h(F)$ in the sparsity argument below.

Let $h\geq1$, and let $f$ be an edge in one copy of $T_N$. Select an internal endpoint of $f$ as the root of this copy
and orient all edges away from the root. After deleting $f$, every internal vertex has exactly two outgoing edges.
Starting from $Q_{h-1}(H)$, add the $N-2$ internal vertices in decreasing order of depth, together with their outgoing
edges. Each added vertex is adjacent to two already existing vertices, so \autoref{lem:rigidity-operations} shows
that its addition increases the rank by two. Hence $r_2(Q_h(H)-f)=r_2(Q_{h-1}(H))+2(N-2)$.

Suppose $h\leq\gamma-1$. For a vertex set $Z$, let $t$ be its
number of terminals in $Z$ and $i_j$ its number of internal vertices in copy
$j$. For each copy, select a terminal outside $Z$ as the root if one exists, and orient all edges away from the root.
Each internal vertex then has two outgoing edges, while every terminal except the root has none. Thus, copy $j$
contributes at most $2i_j$ edges induced by $Z$, with at most one additional edge when $t=N$.

We verify $(2,3)$-sparsity for every $Z$ with $|Z|\geq2$. If
$t\leq1$, the induced graph is a forest, since copies meet in at most
one retained terminal and there is at most one edge at $a$. Thus its
edge count is at most $|Z|-1\leq2|Z|-3$.
Now let $t\geq2$, first with $a\in Z$, and put $i=\sum_j i_j$. If
$t<N$, the terminal forest contributes at most $t-1$ edges, giving
at most $t+(t-1)+2i=2|Z|-3$ edges. If $t=N$, the edge count is at
most $N+(N-\gamma)+2i+h$, which is at most $2N+2i-1=2|Z|-3$ since
$h\leq\gamma-1$. If $a\notin Z$, add $a$ and apply the preceding
bound. Removing its $t\geq2$ incident edges gives at most
$2(|Z|+1)-3-t\leq2|Z|-3$ edges. Hence all edges of $Q_h(F)$ are independent in $\mathcal R_2(Q_h(F))$.
Since $F$ has $N-\gamma$ edges, we obtain
\[r_2(Q_h(H))=r_2(Q_h(F))=|E(Q_h(F))|=2N-\gamma+h(2N-3)=2n-2-\gamma+h.\]

At $h=\gamma-1$ this is $2n-3$, and $Q_h(H)$ is rigid by
\autoref{thm:laman}. The identity $r_2(Q_h(H)-f)=r_2(Q_{h-1}(H))+2(N-2)$ proved
above shows that $Q_h(H)-f$, and hence $Q_h(H)$, is rigid whenever
$Q_{h-1}(H)$ is rigid. Induction on $h$ makes $Q_h(H)$ rigid for
every $h\geq\gamma-1$, proving the first formula. The second formula follows by applying the first at $h-1$ to
$r_2(Q_h(H)-f)=r_2(Q_{h-1}(H))+2(N-2)$, noting that $Q_{h-1}(H)$ has
$n-(N-2)$ vertices.
\end{proof}

\begin{lemma}
\label{lem:tree-global-rigidity}
For $h\geq2$, the graph $Q_h(H)$ from the construction above is globally
rigid if and only if $H$ has at most $h$ connected components.
\end{lemma}

\begin{proof}
\label{proof:tree-global-rigidity}
Write $\gamma=\gamma(H)$.

\emph{Necessity.}
Suppose $Q_h(H)$ is globally rigid. By
\autoref{lem:tree-rigidity}, if $\gamma\geq h+2$,
then $r_2(Q_h(H))<2n-3$, so $Q_h(H)$ is not rigid.
If $\gamma=h+1$, then $Q_h(H)$ is rigid, but
$r_2(Q_h(H)-f)=2n-4$ for every tree edge $f$.
Thus $Q_h(H)$ is not redundantly rigid.
In either case, $Q_h(H)$ cannot be globally rigid
by \autoref{thm:global-rigidity}. So $\gamma\leq h$.

\emph{Sufficiency.}
Suppose $\gamma\leq h$. We first show that $Q_h(H)$
is 3-connected. Let $X$ be any set of at most two
vertices. Consider any copy of $T_N$ after deleting
$X$. We claim that every surviving component contains
a surviving terminal. Otherwise, let $U$ be a component
consisting only of internal vertices. Since every
internal vertex has degree three and $U$ is a tree,
the number of edges joining $U$ to vertices outside
$U$ in the original tree is $3|U|-2(|U|-1)=|U|+2\geq3$.
All these edges must have their other endpoints in
$X$. However, by acyclicity, each vertex outside $U$
has at most one neighbor in $U$, so deleting at most
two vertices cannot remove all these edges, a
contradiction.

Thus every surviving tree component contains a
surviving terminal. If $a\notin X$, all surviving
terminals are connected through $a$. If $a\in X$,
at most one other vertex is deleted. Since $h\geq2$
and the tree copies have disjoint internal vertices,
some copy has no deleted internal vertex and connects
all surviving terminals. Consequently, all surviving
vertices lie in one connected component. Hence
$Q_h(H)$ is 3-connected.

It remains to prove redundant rigidity. Since
$\gamma\leq h$, the two formulas in
\autoref{lem:tree-rigidity} give $r_2(Q_h(H))=r_2(Q_h(H)-f)=2n-3$
for every tree edge $f$. Thus $Q_h(H)$ is rigid,
and no tree edge is a coloop. Choose a rigidity circuit $C$ containing a tree
edge $f$. Such a circuit exists because $f$ is not
a coloop. By
\autoref{lem:rigidity-matroid-components}, each vertex in $C$
has minimum degree at least three and $C$ is redundantly
rigid on $V[C]$. Every internal vertex of the
tree containing $f$ has degree exactly three in
$Q_h(H)$. Hence, whenever such a vertex belongs to
$C$, all three of its incident edges must belong
to $C$. Starting from an internal endpoint of $f$
and repeating this argument along the tree forces
$C$ to contain every edge of that tree, and hence
all terminals.

We now extend the redundantly rigid subgraph $C$.
If $a$ is not already in its support, add $a$
adjacent to all $N\geq4$ terminals. By
\autoref{lem:rigidity-operations}, this preserves
redundant rigidity. Adding all remaining non-tree
edges also preserves redundant rigidity, since
their endpoints are already present. Thus every
non-tree edge belongs to a redundantly rigid
subgraph and is therefore not a coloop. We have shown that no edge of $Q_h(H)$ is a coloop.
Since $Q_h(H)$ is rigid, it is redundantly rigid
by \autoref{lem:rigidity-matroid-components}.
Together with 3-connectivity,
\autoref{thm:global-rigidity} implies that
$Q_h(H)$ is globally rigid.
\end{proof}

\subsection{Weighted edge and vertex deletion}
\label{sec:weighted-deletion}

Let $G$ be a graph on at least four vertices containing a triangle $auv$. Let $G'$ be obtained by deleting the edge $uv$,
adding a new vertex $x$, and joining $x$ to $a,u,v$. We show that $G$ is rigid if and only if $G'$ is rigid,
and likewise for global rigidity.
For rigidity, the triangle $auv$ and the graph on $\{a,u,v,x\}$ with edges $\{au,av,ax,ux,vx\}$ are both rigid
and share the same three attachment vertices $a,u,v$. Thus \autoref{lem:rigidity-substitution} applies in both directions.
For global rigidity, if $G$ is globally rigid, then so is $G'$, since the operation is a \emph{1-extension},
which preserves global rigidity for graphs on at least four vertices \cite{Connelly2005,JacksonJordan2005}.
Conversely, if $G'$ is globally rigid, replacing the graph on $\{a,u,v,x\}$ by the globally rigid triangle
$auv$ gives $G$, which is globally rigid by \autoref{lem:rigidity-substitution}.

Apply the replacement described above to every terminal edge $uv$ of $Q_h(H)$,
using a distinct new vertex for each edge, and denote the resulting graph by $P_h(H)$.
Each replacement is valid because the edges $au$ and $av$ remain present.
By the preceding equivalences, $Q_h(H)$ is rigid if and only if $P_h(H)$ is rigid, and likewise for global rigidity.
Thus, for $N\geq4$ and $h\geq2$, \autoref{lem:tree-rigidity}
and \autoref{lem:tree-global-rigidity} imply that both graphs
are rigid if and only if $\gamma(H)\leq h+1$, and globally
rigid if and only if $\gamma(H)\leq h$.

Take a minimum $k$-cut instance $(H,k,b)$, asking whether deleting at most $b$ edges of $H$ leaves at least $k$
connected components. Assume that $H$ is connected and simple, with $N=|V(H)|$, $m=|E(H)|$, and $4\leq k\leq N$.
Cases with $k<4$ are polynomial-time solvable, while $k>N$ is infeasible, so these cases can be mapped to fixed yes- or
no-instances. For the remaining instances, choose the graph and parameter as follows.

\begin{center}
\begin{tabular}{lccc}
\hline
Property to destroy & Edge deletion & Vertex deletion & Parameter \\
\hline
Rigidity & $Q_h(H)$ & $P_h(H)$ & $h=k-2$ \\
Global rigidity & $Q_h(H)$ & $P_h(H)$ & $h=k-1$ \\
\hline
\end{tabular}
\end{center}

For edge deletion, assign cost one to each terminal edge and cost $m+1$ to each structural edge. For vertex deletion,
assign cost one to each vertex introduced by the replacements and cost $m+1$ to every vertex originally in $Q_h(H)$.
Both choices of $h$ satisfy $h\geq2$. Since $H$ is connected, $\gamma(H)=1\leq h$, so every constructed graph is
globally rigid by the threshold equivalences.

Deleting all $m$ cost-one edges or vertices destroys the target property,
so the optimum cost is at most $m$ and no optimum deletes an object of cost $m+1$. For any $F\subseteq E(H)$,
deleting the corresponding terminal edges from $Q_h(H)$ gives $Q_h(H-F)$, while deleting the corresponding replacement
vertices from $P_h(H)$ gives $P_h(H-F)$. By the threshold equivalences, the target property is destroyed in either
construction exactly when $\gamma(H-F)\geq k$. Therefore, in all four cases, the optimum deletion cost equals
the minimum $k$-cut value of $H$. Keeping the same budget $b$ establishes all four reductions.

Both constructions are simple graphs with $O(N^2)$ vertices and edges.
The larger cost $m+1$ is at most the number of edges in the constructed
graph plus one for edge deletion, and at most the number of vertices
for vertex deletion. This proves \autoref{thm:weighted-destruction}.

\subsection{Unit costs by protecting structural edges}
\label{sec:unit-hardness}

Take the same minimum $k$-cut instance $(H,k,b)$, with
$N=|V(H)|$ and $m=|E(H)|$, and set $h=k-2$.
We construct a graph $\widehat Q(H,k)$ from $Q_h(H)$ as follows.
For each structural edge $uv$ of $Q_h(H)$, introduce $m+1$
new vertices and add all missing edges among these vertices
and the existing vertices $u,v$. The original edge $uv$
is retained, so these $m+3$ vertices form a clique.
The new vertices introduced for different structural edges
are distinct and have no neighbors outside their respective
cliques. All terminal edges remain unchanged.
Denote the resulting simple graph by $\widehat Q(H,k)$.

We first show that a clique on $q\geq3$ vertices remains rigid
after deleting at most $q-3$ edges. We prove this by induction
on $q$. The case $q=3$ is immediate. For $q>3$, the claim
is also immediate if no edge is deleted. Otherwise, choose
a vertex $v$ incident with a deleted edge. After removing $v$,
the remaining $q-1$ vertices have lost at most $q-4$ edges,
so their induced subgraph is rigid by induction. Moreover,
$v$ retains at least $(q-1)-(q-3)=2$ neighbors.
Adding $v$ back therefore preserves rigidity by
\autoref{lem:rigidity-operations}.

Now consider any edge set $D$ of size at most $m$ in
$\widehat Q(H,k)$, and let $F\subseteq E(H)$ be the terminal
edges contained in $D$. Since every inserted clique has
$m+3$ vertices and loses at most $m$ edges, it remains rigid
by the preceding argument.

For each structural edge $uv$, the surviving clique
meets the rest of the graph only at the original vertices
$u,v$. Both this surviving clique and the single edge $uv$
are rigid. Hence, by \autoref{lem:rigidity-substitution},
we may replace the surviving clique by the edge $uv$
without changing the rigidity of the entire graph,
even if $uv$ itself was deleted. Applying this substitution
to every clique leaves exactly $Q_h(H-F)$. Therefore,
$\widehat Q(H,k)-D \text{ is rigid} \Longleftrightarrow Q_h(H-F) \text{ is rigid}$.

Since $H$ is connected, $Q_h(H)$ is rigid by \autoref{lem:tree-rigidity}, and the equivalence above
shows that $\widehat Q(H,k)$ is also rigid. Deleting all $m$ terminal edges leaves $H$ with
$N\geq k$ isolated vertices. Since $h=k-2$,
\autoref{lem:tree-rigidity} implies that this deletion
makes $\widehat Q(H,k)$ nonrigid. Thus an optimum
deletion set has size at most $m$.

Let $D$ be an optimum deletion set, and let $F$ be
its deleted terminal edges. By the equivalence above,
$Q_h(H-F)$ is nonrigid. Applying the same equivalence
when only $F$ is deleted shows that removing the
terminal edges $F$ alone already destroys rigidity.
Hence an optimum cannot contain any edge from an
inserted clique, and every optimum uses only terminal
edges.
For any $F\subseteq E(H)$, \autoref{lem:tree-rigidity}
and $h=k-2$ give $Q_h(H-F)\text{ is nonrigid} \Longleftrightarrow \gamma(H-F)\geq k$.
Consequently, the minimum number of edges whose
deletion makes $\widehat Q(H,k)$ nonrigid equals
the minimum $k$-cut value of $H$. Keeping the same
budget $b$ therefore gives the required reduction.

Finally, the construction produces a simple graph of
polynomial size. Since $h<k\leq N$, the graph $Q_h(H)$
has $N+h(2N-3)=O(N^2)$ structural edges.
Each structural edge introduces $m+1=O(N^2)$ new
vertices and $O(m^2)=O(N^4)$ additional edges.
Thus $\widehat Q(H,k)$ has $O(N^4)$ vertices and
$O(N^6)$ edges. This proves the NP-hardness claim
of \autoref{thm:unit-cost-classification}.

\section*{AI disclosure}
The main result of this paper was discovered by Xean~\cite{Xu2026Xean} with GPT-6~Astra.
The authors wrote the paper with assistance from ChatGPT~6.
Generative AI was thus used both for research discovery and for
preparing the manuscript. The authors take full responsibility for
the correctness and content of the paper, including all material
generated with AI assistance.

\bibliography{main}

@article{Laman1970,
  author  = {Gerard Laman},
  title   = {On Graphs and Rigidity of Plane Skeletal Structures},
  journal = {Journal of Engineering Mathematics},
  volume  = {4},
  number  = {4},
  pages   = {331--340},
  year    = {1970},
  doi     = {10.1007/BF01534980}
}

@article{JacksonJordan2005,
  author  = {Bill Jackson and Tibor Jord{\'a}n},
  title   = {Connected Rigidity Matroids and Unique Realizations of Graphs},
  journal = {Journal of Combinatorial Theory, Series B},
  volume  = {94},
  number  = {1},
  pages   = {1--29},
  year    = {2005},
  doi     = {10.1016/j.jctb.2004.11.002}
}

@article{AsimowRoth1978,
  author  = {Leonard Asimow and Ben Roth},
  title   = {The Rigidity of Graphs},
  journal = {Transactions of the American Mathematical Society},
  volume  = {245},
  pages   = {279--289},
  year    = {1978},
  doi     = {10.1090/S0002-9947-1978-0511410-9}
}

@article{Hendrickson1992,
  author  = {Bruce Hendrickson},
  title   = {Conditions for Unique Graph Realizations},
  journal = {SIAM Journal on Computing},
  volume  = {21},
  number  = {1},
  pages   = {65--84},
  year    = {1992},
  doi     = {10.1137/0221008}
}

@article{JacobsHendrickson1997,
  author  = {Donald J. Jacobs and Bruce Hendrickson},
  title   = {An Algorithm for Two-Dimensional Rigidity Percolation:
             The Pebble Game},
  journal = {Journal of Computational Physics},
  volume  = {137},
  number  = {2},
  pages   = {346--365},
  year    = {1997},
  doi     = {10.1006/jcph.1997.5809}
}

@article{LeeStreinu2008,
  author  = {Audrey Lee and Ileana Streinu},
  title   = {Pebble Game Algorithms and Sparse Graphs},
  journal = {Discrete Mathematics},
  volume  = {308},
  number  = {8},
  pages   = {1425--1437},
  year    = {2008},
  doi     = {10.1016/j.disc.2007.07.104},
  url     = {https://arxiv.org/abs/math/0702129}
}

@book{Oxley2011,
  author    = {James Oxley},
  title     = {Matroid Theory},
  edition   = {2},
  series    = {Oxford Graduate Texts in Mathematics},
  volume    = {21},
  publisher = {Oxford University Press},
  year      = {2011},
  doi       = {10.1093/acprof:oso/9780198566946.001.0001}
}

@misc{CruickshankEtAl2025,
  author       = {James Cruickshank and Bill Jackson and Tibor Jord{\'a}n and
                  Shin-ichi Tanigawa},
  title        = {Rigidity of Graphs and Frameworks: A Matroid Theoretic
                  Approach},
  year         = {2025},
  howpublished = {arXiv:2508.11636v1},
  url          = {https://arxiv.org/abs/2508.11636v1},
  doi          = {10.48550/arXiv.2508.11636}
}

@incollection{JordanWhiteley2017,
  author    = {Tibor Jord{\'a}n and Walter Whiteley},
  title     = {Global Rigidity},
  booktitle = {Handbook of Discrete and Computational Geometry},
  editor    = {Jacob E. Goodman and Joseph O'Rourke and Csaba D. T{\'o}th},
  edition   = {3},
  chapter   = {63},
  pages     = {1661--1694},
  publisher = {Chapman and Hall/CRC},
  year      = {2017},
  doi       = {10.1201/9781315119601},
  url       = {https://www.csun.edu/~ctoth/Handbook/chap63.pdf},
  note      = {DOI refers to the handbook; URL links to the author manuscript}
}

@article{GoldschmidtHochbaum1994,
  author  = {Olivier Goldschmidt and Dorit S. Hochbaum},
  title   = {A Polynomial Algorithm for the {$k$}-cut Problem for Fixed {$k$}},
  journal = {Mathematics of Operations Research},
  volume  = {19},
  number  = {1},
  pages   = {24--37},
  year    = {1994},
  doi     = {10.1287/moor.19.1.24}
}

@inproceedings{Li2019,
  author    = {Jason Li},
  title     = {Faster Minimum {$k$}-cut of a Simple Graph},
  booktitle = {2019 IEEE 60th Annual Symposium on Foundations of Computer Science (FOCS)},
  pages     = {1056--1077},
  publisher = {IEEE},
  year      = {2019},
  doi       = {10.1109/FOCS.2019.00068},
  url       = {https://arxiv.org/abs/1910.02665v1}
}

@article{Marx2006,
  author  = {D{\'a}niel Marx},
  title   = {Parameterized Graph Separation Problems},
  journal = {Theoretical Computer Science},
  volume  = {351},
  number  = {3},
  pages   = {394--406},
  year    = {2006},
  doi     = {10.1016/j.tcs.2005.10.007}
}

@misc{Razgon2010,
  author       = {Igor Razgon},
  title        = {Computing Multiway Cut within the Given Excess over the
                  Largest Minimum Isolating Cut},
  year         = {2010},
  howpublished = {arXiv:1011.6267v1},
  url          = {https://arxiv.org/abs/1011.6267v1},
  doi          = {10.48550/arXiv.1011.6267}
}

@misc{EGRESDestroyingRigidity,
  author       = {{EGRES Open}},
  title        = {Destroying Rigidity},
  howpublished = {EGRES Open problem page},
  url          = {https://lemon.cs.elte.hu/egres/open/Destroying_rigidity},
  note         = {Revision 1817, accessed October 12, 2026}
}

@article{MotevallianYuAnderson2015,
  author  = {S. Alireza Motevallian and Changbin Yu and Brian D. O. Anderson},
  title   = {On the Robustness to Multiple Agent Losses in {2D} and {3D} Formations},
  journal = {International Journal of Robust and Nonlinear Control},
  volume  = {25},
  number  = {11},
  pages   = {1654--1687},
  year    = {2015},
  doi     = {10.1002/rnc.3167}
}

@article{Jordan2021,
  author  = {Tibor Jord{\'a}n},
  title   = {Minimum Size Highly Redundantly Rigid Graphs in the Plane},
  journal = {Graphs and Combinatorics},
  volume  = {37},
  number  = {4},
  pages   = {1415--1431},
  year    = {2021},
  doi     = {10.1007/s00373-021-02327-4}
}

@article{EdmondsKarp1972,
  author  = {Jack Edmonds and Richard M. Karp},
  title   = {Theoretical Improvements in Algorithmic Efficiency for Network Flow Problems},
  journal = {Journal of the ACM},
  volume  = {19},
  number  = {2},
  pages   = {248--264},
  year    = {1972},
  doi     = {10.1145/321694.321699}
}

@article{Connelly2005,
  author  = {Connelly, Robert},
  title   = {Generic Global Rigidity},
  journal = {Discrete \& Computational Geometry},
  volume  = {33},
  number  = {4},
  pages   = {549--563},
  year    = {2005},
  doi     = {10.1007/s00454-004-1124-4}
}

@article{HopcroftTarjan1973,
  author  = {John E. Hopcroft and Robert E. Tarjan},
  title   = {Dividing a Graph into Triconnected Components},
  journal = {SIAM Journal on Computing},
  volume  = {2},
  number  = {3},
  pages   = {135--158},
  year    = {1973},
  doi     = {10.1137/0202012}
}

@inproceedings{VanDenBrandEtAl2023,
  author    = {Jan van den Brand and Li Chen and Rasmus Kyng and Yang P. Liu and
               Richard Peng and Maximilian {Probst Gutenberg} and
               Sushant Sachdeva and Aaron Sidford},
  title     = {A Deterministic Almost-Linear Time Algorithm for Minimum-Cost Flow},
  booktitle = {2023 IEEE 64th Annual Symposium on Foundations of Computer Science (FOCS)},
  pages     = {503--514},
  publisher = {IEEE},
  year      = {2023},
  doi       = {10.1109/FOCS57990.2023.00037},
  url       = {https://arxiv.org/abs/2309.16629}
}

@misc{Xu2026Xean,
  author       = {Chao Xu},
  title        = {{Xean}},
  year         = {2026},
  howpublished = {Software repository},
  url          = {https://github.com/chaoxu/xean}
}

@book{GraverServatiusServatius1993,
  author    = {Jack E. Graver and Brigitte Servatius and Herman Servatius},
  title     = {Combinatorial Rigidity},
  series    = {Graduate Studies in Mathematics},
  volume    = {2},
  publisher = {American Mathematical Society},
  year      = {1993},
  doi       = {10.1090/gsm/002}
}

@article{AspnesEtAl2006,
  author  = {James Aspnes and Tolga Eren and David K. Goldenberg and
             A. Stephen Morse and Walter Whiteley and Yang Richard Yang and
             Brian D. O. Anderson and Peter N. Belhumeur},
  title   = {A Theory of Network Localization},
  journal = {IEEE Transactions on Mobile Computing},
  volume  = {5},
  number  = {12},
  pages   = {1663--1678},
  year    = {2006},
  doi     = {10.1109/TMC.2006.174}
}

@article{YuAnderson2009,
  author  = {Changbin Yu and Brian D. O. Anderson},
  title   = {Development of Redundant Rigidity Theory for Formation Control},
  journal = {International Journal of Robust and Nonlinear Control},
  volume  = {19},
  number  = {13},
  pages   = {1427--1446},
  year    = {2009},
  doi     = {10.1002/rnc.1386}
}
\end{document}